# > 2π Phase Modulation using Exciton-Polaritons in a Two-Dimensional Superlattice

Jason Lynch[1], Pawan Kumar[1, 2], Chen Chen[3], Nicholas Trainor[3, 4], Shalina Kumari[3, 4], Tzu-Yu Peng[5, 6], Cindy Yueli Chen[7], Yu-Jung Lu[5, 6], Joan Redwing[3, 4], Deep Jariwala[1, *]

[1] Electrical and Systems Engineering, University of Pennsylvania, Philadelphia, PA 19104

[2] Inter-University Microelectronics Center, Leuven 3001, Belgium

[3] 2D Crystal Consortium Materials Innovation Platform, Materials Research Institute, Penn State University, University Park, PA, USA

[4] Materials Science and Engineering, Penn State University, University Park, PA, USA

[5] Research Center for Applied Sciences, Academia Sinica, Taipei 11529, Taiwan

[6] Graduate Institute of Applied Physics, National Taiwan University, Taipei 10617, Taiwan

[7] Department of Chemistry, University of Pennsylvania, Philadelphia, PA 19104, USA

*Corresponding author: dmj@seas.upenn.edu

## Abstract

Active metamaterials promise to enable arbitrary, temporal control over the propagation of wavefronts of light for applications such as beam steering, optical communication modulators, and holograms. This has been done in the past using patterned silicon photonics to locally control the phase of light such that the metasurface acts as a large number of wavelets. Although phase modulation only requires refractive index modulation when the interaction length is on the order of the wavelength, this is not enough to significantly modulate the phase of light in flatland. Instead, phase modulation is achieved using a resonant mode such as a plasmon or high-Q cavity mode that enable light to accumulate a large amount of phase over a short distance and coupling it to an active material that modulates the light-matter interactions. Here, we report that electrostatic doping can modulate the light-matter interaction strength of a two-dimensional $WS_2$ based multi quantum well (MQW) structure going from strongly-coupled, phase-accumulating exciton-polaritons to weakly-coupled exciton-trion-polaritons. As a result of this transition, 2.02π radians of phase modulation is observed using spectroscopic ellipsometry. This result demonstrates the potential of the MQW structure as a compact, lightweight electro-optical modulators for LiDAR and optical communications in the red region of visible spectrum.

## Introduction

As one of the quintessential classes of two-dimensional (2D) semiconductors, transition metal dichalcogenides (TMDCs) of Mo and W have shown the promise to facilitate flat-optical systems[1,2] such as beam steering[3], metalenses[4], and electro-optical modulators[5,6]. Additionally, their large band gaps are ideal for visible light communications (VLC)[7] and holographics[8]. The direct band gap nature of TMDCs in the monolayer limit[9], and their large refractive index[10], are ideal for efficient, subwavelength photonics applications. Additionally, their strong excitons (Coulomb bound electron-hole pairs) in the visible and near infrared ranges[11] host strong-light matter interactions. The strength of these excitons depends heavily on the binding energy of the constituent electron and hole pairs[12], and therefore, the light-matter interactions within TMDCs can be modulated through the binding energy[13,14]. Through the tuning of the exciton binding energy, multiple TMDCs have shown highly gate-tunable, complex refractive index ($\tilde{n} = n + ik$) values[15–17] which is highly sought after in electro-optics.

In contrast to classical phase modulators, such as those based on Si[18] and III-V semiconductors[19], which modulate their optical phase length to control the phase, basic TMDC-based phase modulators rely on tuning bare excitons and trions to modulate the amplitude and phase of reflected light. Recently, phase modulation of incident light up to π/5 radians has been demonstrated in monolayer TMDCs purely using the exciton resonance and the relatively slow ion gel gating method[20]. However, a single resonance cannot induce a phase shift > π. Instead, a second resonance is required for full 2π phase modulation of light which is vital for applications such as phased arrays. It is common to do this by coupling excitons to a cavity mode to form part-matter, part-light quasiparticles called exciton-polaritons[21]. In this case, the cavity mode confines the light and causes it to accumulate a large degree of phase in a small volume while the exciton provides tunability. Excitons have been coupled to waveguide modes in the past[22,23], but previous phase modulators using TMDCs have focused on coupling to plasmon-polaritons [6,24–26]. However, the use of plasmonics increases the loss of the system, and the exciton tunability of should decrease along with its binding energy when it is placed in close proximity with a metal[27]. In our previous work, we demonstrated a multi quantum well (MQW) or superlattice of $WS_2$ separated by insulators on top of a reflective substrate to form strongly coupled exciton-polaritons[28]. The insulating layers electronically separated the $WS_2$ monolayers allowing them to keep the direct band gap characteristics including the large exciton binding energy[29] while increasing the light-matter interactions. Since the TMDCs maintain their large binding energy, they also maintain their high tunability which can be leveraged to produce highly efficient, compact phase modulators using exciton polaritons.

In this work, we demonstrate that exciton-polaritons enable full 360° phase modulation in a superlattice of $WS_2$ and $Al_2O_3$ on a $SiO_2$/Si substrate. Instead of forming an external cavity using top and bottom reflectors or nanopatterning, self-hybridized, strongly-coupled exciton-polaritons form under transverse electric (TE) polarized light at an angle of incidence (AOI) of 55°. The light-matter interaction strength is tuned using electrostatic doping, and the exciton and trion coupling parameters are modulated by -23% and +129%, respectively. Leveraging the fact that the phase of transverse magnetic (TM) polarized light is unaffected by electrostatic doping, spectroscopic ellipsometry is used to experimentally verify that 2.02π phase modulation is occurring. Although the loss of the phase modulator remains large, it achieves full control of the phase of light which is vital for LiDAR and free-space optical communications.

## Results and Discussion

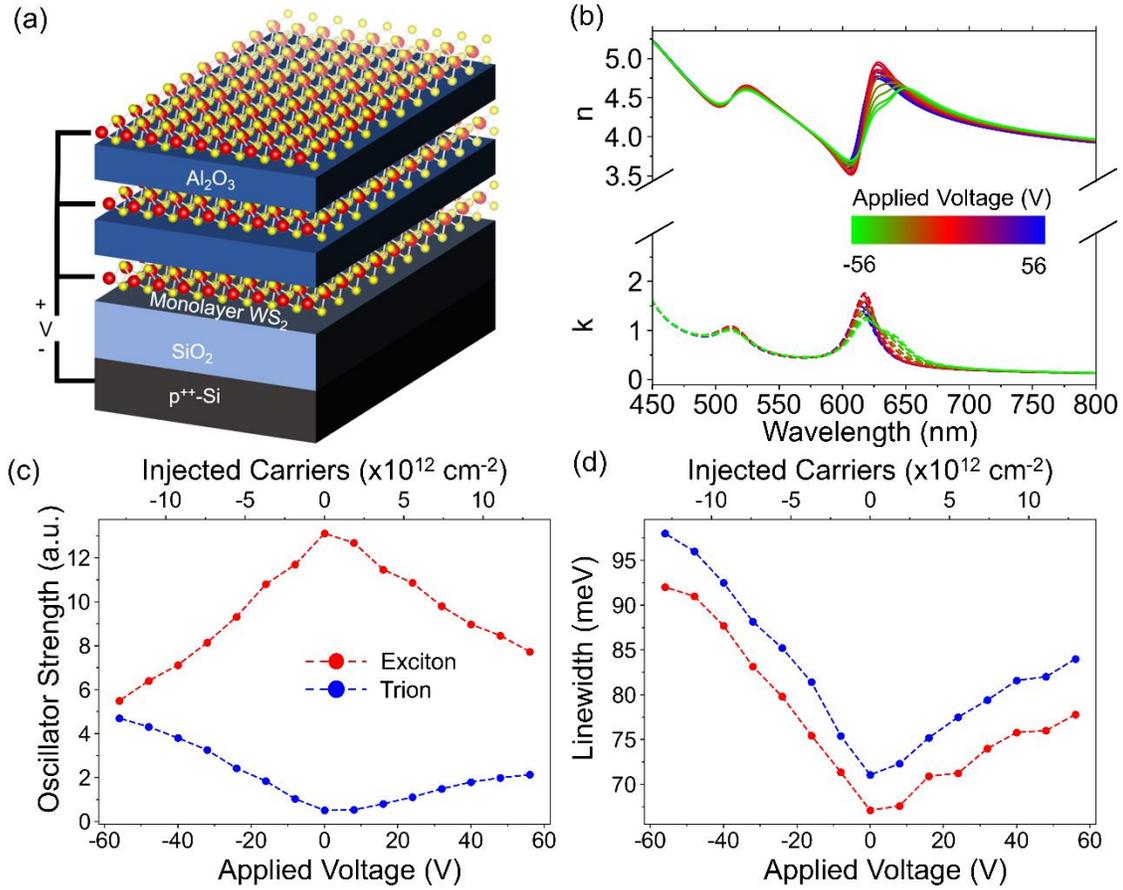

**Figure 1. Gate-tunable optical properties of monolayer WS$_2$.** **(a)** Schematic of the phase tuning superlattice showing its repeated geometry and electrical contact to the WS$_2$ layers. **(b)** The real (solid lines) and imaginary (dashed) parts of the gate-dependent refractive index measured using spectroscopic ellipsometry and voltages ranging from -56 V to 56 V (8 V steps). The extracted gate-dependent **(c)** oscillator strength and **(d)** linewidth of the exciton and trion are from the Lorentz model. The injected carriers (N) is calculated using a parallel plate capacitor model ($N = \frac{\varepsilon_{SiO2}\varepsilon_0 V}{ed}$ where t$_{SiO2}$ is the SiO$_2$ thickness (93 nm), $\varepsilon_{SiO2}$ is the permittivity of SiO$_2$ (3.9), e is the elementary charge, and $\varepsilon_0$ is the vacuum permittivity). The schematic in (a) was prepared using the VESTA software[30].

Large-area monolayers of WS$_2$ are grown using a metal-organic chemical vapor deposition (MOCVD) technique which has been shown to produce films with excellent electrical properties[31,32]. A wet transfer technique is used to transfer centimeter-sized monolayers on to the SiO$_2$/Si, and atomic layer deposition (ALD) is used to deposit the Al$_2$O$_3$ insulating layers to fabricate the superlattice (See Methods). The wet transfer technique has been shown to maintain the high electrical quality of the monolayers[31,33]. The electrical contact (10 nm of Ti and 40 nm of Au) is deposited to make electrical contact to the bottom most WS$_2$ layer. The unit cell of monolayer WS$_2$ and 3 nm of Al$_2$O$_3$ is repeated 3 times to form the superlattice (Figure 1a). All three WS$_2$ layers are connected to the same voltage while the bottom most layer has the most charge injected into it due to the Debye length of WS$_2$ and ALD-grown Al$_2$O$_3$ are on the scale of a few nanometers[34] and tens of nanometers[35], respectively. The superlattice has been shown to maintain the direct band gap nature of the constitute monolayers[28], and an N = 3 structure is chosen to enhance the light-matter coupling while maintaining a large degree of tunability by modulating a single monolayer (Supporting Information Figure S2).

The gate-tunable, complex refractive index of monolayer WS$_2$ on an SiO$_2$ (93 nm)/p$^{++}$-Si substrate is measured using spectroscopic ellipsometry to accurately model the optical properties of the sample (See Methods and Supporting Information). The ellipsometry data is fitted to a multi-Lorentz oscillator model to extract the complex refractive index (Figure 1b). The calculated values match well with the experimental data (Supporting Information Figure S3). Additionally, the tunability of the monolayer is confirmed using gate-dependent photoluminescence (PL) (Supporting Information Figure S4). The optical properties of WS$_2$ are modulated by injecting charge with an applied voltage to the metal-oxide-semiconductor Capacitor (MOSCap) geometry. The voltage is applied to the WS$_2$ layer while the p$^{++}$-Si substrate is grounded so a positive (negative) applied voltage injects holes (electrons) into the WS$_2$. The injected free charge modulates the Coulomb screening of the excitons that dominate the optical properties of WS$_2$. The increased screening decreases the oscillator strength and increases the damping of the exciton states (Figures 1c and 1d). However, the free carriers can also bind to the neutral excitons to form charged triplet states (trions) which are seen as a new resonance that is slightly red-shifted from the exciton. The charge of the injected carriers determines the charge of trions. Interestingly, the majority of the oscillator strength lost by the exciton is gained by the trion consistent with previous studies[15,20,36]. The trion peak is only observed when electrons are injected (V < 0 V) because the trion redshifts away from the exciton peak allowing it to be partially resolved, but not fully resolved. However, when holes are injected, the trion blueshifts towards the exciton preventing its direct observation[13,15,20]. The injected carrier concentration (N) has been calculated using a parallel plate capacitor model (N = $\frac{\varepsilon_{SiO2}\varepsilon_0 V}{ed}$ where t$_{SiO2}$ is the SiO$_2$ thickness (93 nm), $\varepsilon_{SiO2}$ is the permittivity of SiO$_2$ (3.9), e is the elementary charge, and $\varepsilon_0$ is the vacuum permittivity). Note that the injected carrier concentration differs from the free carrier concentration since WS$_2$ is n-type under ambient conditions. The zero point for the free carrier concentration occurs between 0 and 8 V as seen by the relatively weak modulation of the oscillator strength in this range.

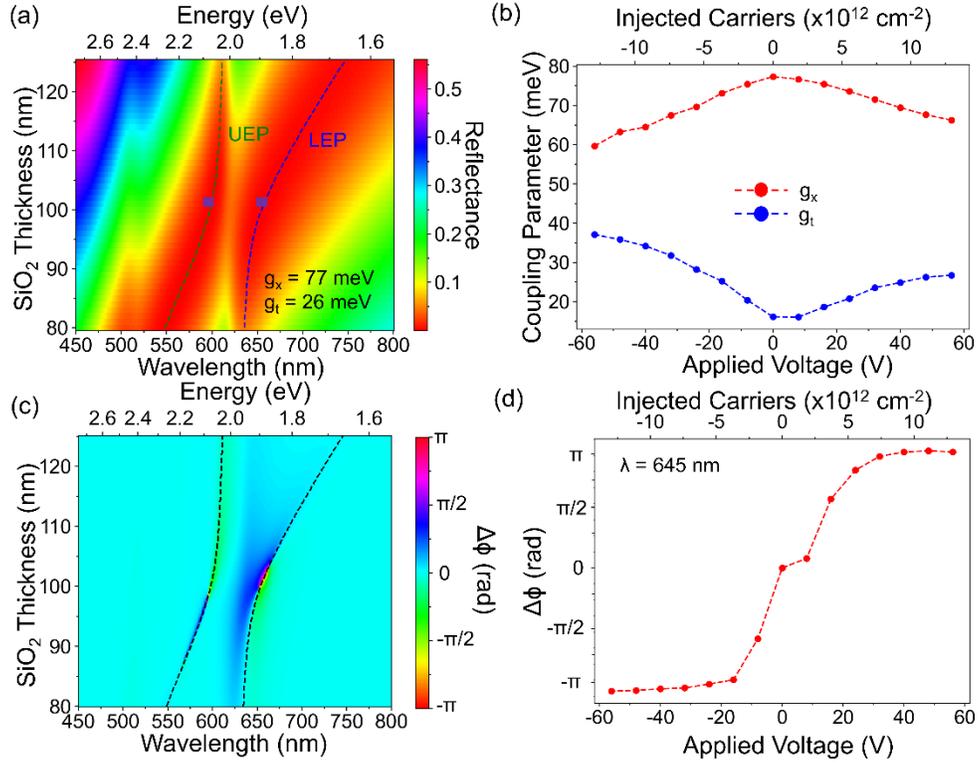

**Figure 2. Simulated modulation of exciton-polaritons. (a)** The simulated dispersion of TE polarized light without an applied voltage. The dispersion shows the anticrossing behavior that is the defining featuring of excitons and cavity modes hybridizing into polaritons. The experimentally measured UEP and LEP wavelength of our superlattice is overlaid (Purple boxes), and they agree well with theory. The upper (UEP) and lower (LEP) exciton-polariton branches are fitted to a three-coupled-oscillator model to calculate the exciton ($g_x$) and trion ($g_t$) coupling parameters. **(b)** The gate-dependent coupling parameters show the same trend as the oscillator strengths since $g \propto \sqrt{\frac{f}{V_m}}$. **(c)** The modulated phase dispersion is found to closely follow the UEP and LEP demonstrating that the phase modulation is driven by the presence of the polaritons. **(d)** The gate-dependent phase modulation at a single wavelength ($\lambda$ = 652 nm). The phase is modulated by 2.07$\pi$ radians over the range of -56 V to 56 V showing that the superlattice can fully control the phase of reflected light.

When an exciton strongly couples to a cavity mode by energy being exchanged between the two states more rapidly than it is dissipated to the environment, the two states hybridize to form exciton-polaritons. Exciton-polaritons by nature confine light to a small volume due to their cavity origins and have a large degree of tunability due to their excitonic origins. This simple 2-mode system is modelled using the Jaynes-Cummings model[37]. However, when a third mode is present, such as trions in our case, a three-coupled oscillator model is required where the cavity mode couples to both the exciton and trion (See Supporting Information)[38]. The strength of the light-matter interactions is captured by the coupling parameter (g), and it depends on both the oscillator strength (f) of the matter resonances and the mode volume of the cavity ($g \propto \sqrt{\frac{f}{V_m}}$)[39]. A single monolayer of WS$_2$ typically produces excitonic coupling parameters ($g_x$) in the range of approximately 10 to 30 meV depending on the cavity mode at room temperature[40–42]. However, using the transfer matrix method (See Methods) to simulate the dispersion of our system, our system hosts an exciton coupling parameter of $g_x$ = 78 meV and a trion coupling parameter of $g_t$ = 16 meV without an applied voltage and a 55° AOI for TE polarized light (Figure 2a). The TMM is found to agree with the experimental superlattice demonstrating its accuracy (Supporting Information Figure S5). The exciton coupling parameter can be further increased to 96 meV at an incident angle of 80°, but this comes at the expense of tunability (Supporting Information Figure S6). The injection of charge in the bottom most layer is found to alter the dispersion of the system by both modulating the coupling parameter, through the modulation of the oscillator strengths, and by modulating the trion energy. The excitonic coupling parameter is found to decrease by 23% and $g_t$ increases by 129% by injecting 1.3 x $10^{13}$ electrons/cm$^2$ in the WS$_2$ (Figure 2b). The reduction is large enough that the system is no longer in the strong light-matter coupling regime as evident by the lack of an exciton transparency when a voltage of -56 V is applied. Therefore, by injecting charge, the system goes from a hybridized state with exciton-polaritons to an unhybridized one with a pure exciton and pure cavity mode. Although a three-coupled-oscillator should produce three polariton modes, only two polariton peaks are observed because the trion and exciton peaks are not fully resolved from one another. This results in two of the polariton modes are convoluted into one[43].

      Since exciton-polaritons are the byproduct of strong light-matter interactions, they are excellent at confining light. Therefore, the phase of reflected light is heavily influenced by the presence of exciton-polaritons. Because of this, modulating exciton-polaritons is a highly-efficient method for modulating phase. This is seen in the simulated phase modulation as its dispersion is found to closely follow the dispersion of the unbiased superlattice (Figure 2c). Therefore, exciton-polaritons can easily modulate the phase of reflected light as shown in the gate-dependent phase modulation at λ = 645 nm where > 2π phase modulation is predicted (Figure 2d). When electrons are injected into the WS$_2$, the exciton oscillator strength drops significantly, and the system goes from the hosting hybridized polaritons to weakly-coupled excitons, trions, and a cavity mode. In this case, the polariton state no longer occurs below the optical band gap of WS$_2$, although there is still absorption due to the finite linewidth of the unperturbed modes, and the reflected light accumulates less phase. When holes are injected into the WS$_2$, the oscillator strength decreases, but not to the degree to enter the weak coupling regime, causing the polariton to blueshift. In this case, more phase is accumulated at wavelengths slightly below the unbiased polariton wavelength. This is why the wavelength with > 2π is 645 nm compared to the unbiased polariton wavelength (655 nm).

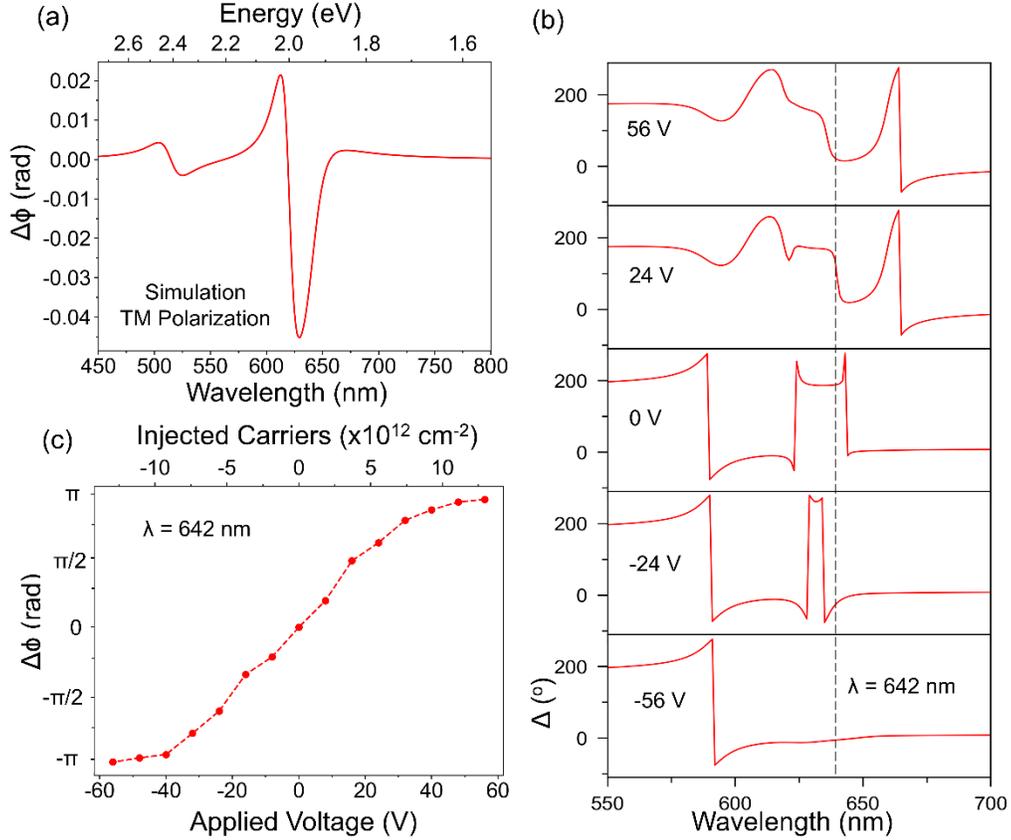

**Figure 3. Experimental phase modulation.** (a) The simulated phase modulation for TM polarized light in our superlattice. Since exciton-polaritons are not excited for TM polarized light, the modulation is purely excitonic, and it is limited since light is not highly-confined in the superlattice by a cavity mode. (b) The experimentally measured Δ (Δ = ϕ$_{TM}$ − ϕ$_{TE}$ where ϕ$_{TM}$ (ϕ$_{TE}$) is the phase of TM (TE) reflected light) using a spectroscopic ellipsometer with an AOI of 55°. Δ is defined over the range of -100° to 260° by convention. Therefore, the sharp decreases are the result of the cyclic nature of phase. (c) The gate-dependent phase modulation for TE polarized light at λ = 642 nm showing full 2π modulation.

An N = 3 superlattice is fabricated to validate the predictions of our simulations (See Methods). We confirmed that the WS$_2$ maintained their high-quality monolayer properties upon stacking using Raman[44] and gate-dependent PL spectroscopy and the presence of exciton-polaritons using angle-dependent reflectance spectroscopy (Supporting Information Figure S7 and S8). Typical phase modulation measurements use a Mach-Zehnder geometry to measure phase modulation in one arm relative to another unmodulated one. However, we leverage the fact that polaritons only form under TE polarized light. Therefore, the phase of TM light should be insensitive to gate voltage. This is validated by simulations that show that the phase modulation of TM light is 0.02 radians at λ = 642 nm over the voltage range studied here (Figure 3a). We then treat the phase of TM light as constant analogous to the reference arm of a Mach-Zehnder interferometer. The spectroscopic ellipsometer measures the phase difference between TE (ϕ$_{TE}$) and TM (ϕ$_{TM}$) light (Δ = ϕ$_{TM}$ − ϕ$_{TE}$) enabling the measurement of phase modulation of TE light (Figure 3b). Note that Δ is defined on the range of -100° to 260° by convention so the sharp decreases in Figure 3b are due to the cyclic nature of phase as opposed to a physical phenomenon. When the system is in the strong coupling region (V > -32 V), the lower exciton-polariton is present below the band gap of $WS_2$. This is seen as Δ highly-dispersive below the exciton wavelength for all voltages except V = -56 V. However, at V = -56 V, there is no sub-gap resonance present, and as a result, Δ is relatively flat in this range. In this case, the closest resonance is the trion, and it will dominate the optical characteristics of the superlattice. Since the trion is purely a matter state, less phase will be accumulated by incident light. As the voltage increases at λ = 642 nm, the cavity mode blueshifts to the operating wavelength due to the weakened total oscillator strength of the exciton-trion states, and Δ decreases because of an increase in $φ_{TE}$ (Δ = $φ_{TM}$ − $φ_{TE}$). This effect is strong enough for 364° phase modulation to occur at 642 nm (Figure 2c). This full 2π phase modulation requires a larger voltage than predicted by theory (Figure 2d), but this is attributed to the deposition of $Al_2O_3$ on top of the monolayer reducing the unbiased exciton binding energy[12,15]. Therefore, the tunability of the monolayer is reduced by the presence of the $Al_2O_3$. Additionally, the required charge injection and electric field strength is comparable to other phase modulators[21,45]. The use of exciton-polaritons also shows a 10x increase in phase modulation using less than 1/4th of the injected carriers of a purely excitonic $WS_2$ system[20]. This shows that polaritons can be leveraged to greatly enhance the efficiency of electro-optical devices.

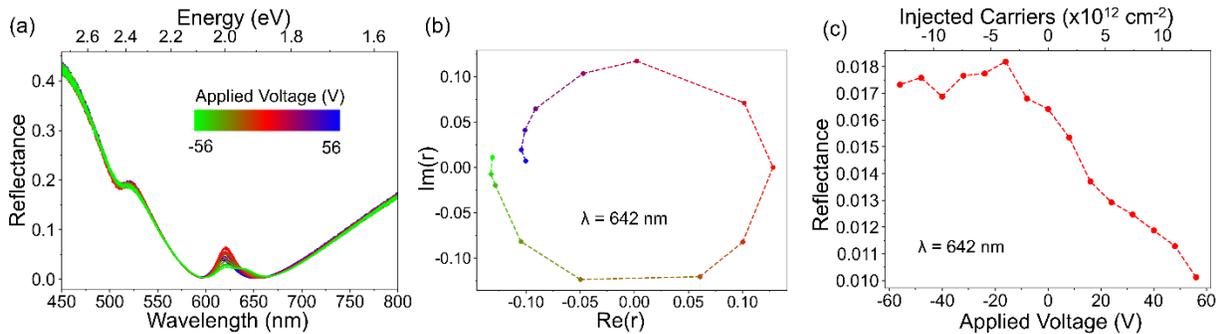

**Figure 4. Loss in the superlattice phase modulator. (a)** Gate-dependent reflectance spectra with an AOI of 55° for TE polarized light. The spectra were measured using both J.A. Woolam M-2000 and W-Vase ellipsometers (See Methods). The **(b)** reflection coefficient phasor (r) and **(c)** gate-dependent reflectance at λ = 642 nm showing the performance of the superlattice at its operating wavelength. The amplitude of the reflection coefficient phasor is measured by taken the square root of the reflectance, and its phase is measured using the gate-dependent Δ values (the phase at V = 0 V is set to 0). The gate-dependent reflectance shows that the superlattice has significant loss (17.8 dB), and the loss decreases (increases) as electrons (holes) are injected into the system.

In addition to max phase modulation, the loss of the system is a key metric for characterizing an electro-optical phase modulator. The gate-dependent reflectance shows the modulation of the spectra, and the largest tunability is seen at the lower polariton wavelength (Figure 4a). The reflectance spectra shows two polariton modes and a reflectance maximum (absorptance minimum) at the exciton wavelength experimentally confirming the presence of exciton-polaritons. The lower polariton also shows more tunability than the upper polariton and the B exciton (510 nm) which is why it is the only resonance where we observe full 2π modulation.

The reflection coefficient phasor is calculated using the reflectance and Δ measurements (Figure 4b). Along with the gate-dependent reflectance at the operating wavelength (Figure 4c), the reflectance is found to vary by 0.008. The reflectance is relatively constant for negative voltages since the decreasing oscillator strength and red-shifting trion energy counteract one another. However, the reflectance drops significantly for positive voltages as the polariton blue shifts towards the operation wavelength and increases the absorptance of the system. The average loss is found to be 17.8 dB. The overall performance is quantified using the well-accepted figure of merit (FoM) of phase modulation per dB loss. This yields a FoM of 0.36 rad/dB which is comparable to other free-space phase modulators.

Table 1 compares the performance of our superlattice with other free-space phase modulators in the visible and near infrared (λ < 1,000 nm). The FoM is found to vary from approximately 0.035 to 0.87 rad/dB for these modulators, and our system is near the middle with a FoM of 0.36 rad/dB. Only two of the modulators are capable of generating < π phase modulation, and both of these systems rely on bulky (millimeter-thick) media[46,47]. Therefore, they require significantly larger volumes of media to be modulated which significantly increases their switching energy and limits their switching speed. Additionally, they use slow modulation methods such as thermos-optical effects that require the thermal energy to be dissipated before the next bit of information can be communicated[47,48], or the physical rotation of the molecules in liquid crystals[46]. Although both of these systems are excellent phase modulators, the use of sub-wavelength active media promises to decrease the switching energy, increase the switching speed, and allow for the integration of phase modulators into ultra-compact devices. Subwavelength phase modulators have also shown excellent performance using non-electrostatic tuning mechanisms. However, they have produced FoM's < 0.3[49,50]. However, these systems are still 100's of nanometers thick, and they do not fully unlock the advantages of the atomic limit.

At the atomic limit, phase modulators have consisted of either graphene coupled to a plasmonic metasurface[51] or planar TMDCs[17,20]. All of these phase modulators rely on a single resonance so their limited to < π phase modulation. Graphene has been excellent for electro-optics at telecommunications wavelengths since it has a highly tunable refractive index. However, the reliance on free-carrier effects is less effective at shorter wavelengths since the Drude model depends less on the free-carrier concentration at higher energies[52]. This is why an injected carrier concentration 7.94 x $10^{13}$ $cm^{-2}$ is needed to produce 5π/9 radians of phase modulation at λ = 760 nm while only 2.3 x $10^{12}$ $cm^{-2}$ is needed to produce a comparable change at λ = 8.5 μm. As for TMDC-based modulators, $MoSe_2$ has been shown to modulate phase by 0.14π [17]. However, this was done at T = 4 K which is undesirable for many electro-optical applications, and the performance rapidly decreased as temperature increased. A FoM of 51 °/dB was achieved in a purely excitonic system of $WS_2$ on an $Al_2O_3$/Al substrate which outperforms our phase modulator by a factor of 2. However, this system used higher-quality, exfoliated monolayers, and an ion gel to inject > 4 times the number of free-carriers. Therefore, a direct comparison between the two modulators to observe the performance enhancement using exciton-polaritons over bare excitons is unreasonable. Further, ion gels rely of physical movement of massive ions for gating (doping) and hence cannot respond to high frequencies of operation[53,54]. To properly compare the two system, we have re-optimized the $WS_2$ on $Al_2O_3$/Al phase modulator using our gate-tunable refractive index of $WS_2$ (Supporting Information Figure S9). We find that the excitonic system gives a phase modulation of 0.028π and a figure of merit of 0.012 rad/dB. Therefore, the use of exciton-polaritons increased the phase modulation by a factor 71 and the FoM by a factor of 10 compared to bare excitons when normalized for WS$_2$ quality and carrier injection.

**Table 1. Comparison of phase modulators in the visible and near infrared (< 1,000 nm).** Comparison of the tuning mechanism, operation wavelength, injected carriers, patterning, phase modulation, and loss of several systems in literature and our work. All of the results are experimental unless otherwise indicated. MQW is a multi-quantum-well, and DBR is a distributed Bragg reflector.

| System | Tuning Mechanism | Operation Wavelength (nm) | Injected Carriers (x10$^{12}$ cm$^{-2}$) | Patterned? | Max Phase Modulation (rad) | Loss (dB) | FoM (rad/dB) |
|---|---|---|---|---|---|---|---|
| Graphene and Au Metasurface[51] | Electrostatic (Free-Carriers) | 760 | 79.4 | Yes | 0.56π | 4.19 | 0.42 |
| Al$_{0.9}$Ga$_{0.1}$As/GaAs MQW on a DBR[49] | Quantum-Confined Stark Effect | 917 | - | Yes | ≈0.04π | 9.51 | ≈0.13 |
| Liquid Crystals between Dielectric Mirrors[46] | Liquid Crystal Rotation | 532 | - | No | 3π | 0.34 | 27.7 |
| LiNbO$_3$ Metasurface (Simulated)[50] | Pockels Effect (Free-Carriers) | 667 | - | Yes | 0.84π | 7.7 | 0.34 |
| Perovskite (MAPBCl$_3$)[47] | Thermo-Optic | 650 | - | No | ≈π | ≈10* | ≈0.31 |
| hBN-Encapsulated MoSe$_2$ on Graphene/SiO$_2$/Si at T = 4 K[17] | Electrostatic (Exciton) | 754 | 9 | No | 0.14π | 12.1 | 0.036 |
| WS$_2$ on Al$_2$O$_3$/Al[20] | Ion Gel (Exciton) | 605 | 54 | No | π/5 | 0.7 | 0.9 |
| Our Work | Electrostatic (Exciton-Polariton) | 642 | 13 | No | 2.02π | 17.8 | 0.36 |

*Loss is simulated using the reported refractive index.

## Conclusion

We have demonstrated that exciton-polaritons in a WS$_2$ superlattice enable > 360° phase modulation by electrostatic doping. The enhanced light-matter interactions of the superlattice compared to a single monolayer creates exciton-polaritons with a coupling parameter of 78 meV without the need of a top reflective layer or nanopatterning. The injection of charge into one of the monolayers in the superlattice modulates the exciton-polaritons by decreasing the excitons oscillator strength and increasing its linewidth while also converting the system to a weakly coupled exciton-trion-polaritons. The exciton and trion coupling parameters are found to change by -23% and +129%, respectively. This process results in full 2π phase modulation of light which is verified experimentally using spectroscopic ellipsometry. Although the system still suffers from loss, its FoM is comparable to similar phase modulators. Additionally, when normalizing for crystal quality and carrier injection, we find that the use of exciton polaritons increase the phase modulation by a factor of 71 and the FoM by a factor of 10 over a similar bare-exciton-based phase modulator. This work serves as aa demonstration that exciton-polaritons in a $WS_2$ superlattice can significantly control the properties of light in the ultrathin regime.

**Methods**

*Fabrication Process*

The growth of monolayer $WS_2$ on 2" diameter c-plane sapphire was carried out in a metalorganic chemical vapor deposition (MOCVD) system equipped with a cold-wall horizontal reactor with an inductively heated graphite susceptor with gas-foil wafer rotation (DOI: 10.60551/znh3-mj13). Tungsten hexacarbonyl ($W(CO)_6$) was used as the metal precursor while hydrogen sulfide ($H_2S$) was the chalcogen source with $H_2$ as the carrier gas. The $W(CO)_6$ powder was maintained at 25 °C and 400 Torr in a stainless-steel bubbler. C-plane sapphire (2" diameter) with a nominal ±0.2° miscut towards the M-axis was used for the growths. The synthesis of $WS_2$ monolayer is based on a multi-step process, consisting of nucleation, ripening, and lateral growth steps, which was described previously[31,32]. In general, the $WS_2$ was nucleated for 30 sec at 850°C, then ripened for 20 min at 850°C and 10 min at 1000°C, and then grown for 20 min at 1000 °C, which gives rise to a coalesced monolayer across the entire 2″ wafer. During the lateral growth, the tungsten flow rate was set as $5.7×10^{-4}$ sccm and the chalcogen flow rate was set as 400 sccm while the reactor pressure was kept at 50 Torr. After growth, the substrate was cooled in $H_2S$ to 300 °C to inhibit the decomposition of the deposited $WS_2$ films. Using this condition, the growth of a fully coalesced monolayer $WS_2$ was achieved across the 2" sapphire substrate. The sample's detailed growth recipe and all characterization data is available at: https://data.2dccmip.org/zGG8RNdCYVgs.

After growth, the 2"-wafers are cleaved into 1 cm x 1cm squares for the transfer process. Poly Methyl Metha Acrylate (PMMA) 950k A4 is spin coated on top of the samples at 2,500 rpm for 45 seconds, and the sample is allowed to air dry. Next, the overhanging PMMA is cleared from the edges using a razor blades, and the sample is submerged in 85 °C de-ionized (DI) water until bubbles form (20-60 minutes). The sample is then placed on top of a 3 M solution of KOH at 85 °C until the film start to delaminate from the sapphire substrate. The sample is then manually dipped with a 45° inclination into the solution to separate the film and the substrate. The floating, PMMA-supported $WS_2$ monolayer is then scooped up by a glass slide and transferred to float on DI water for cleaning. The film is allowed to float for 10 minutes before it is transferred to a fresh DI water container. This step is repeated once more to ensure that the sample is adequately cleaned. The film is then scooped onto its desired substrate and allowed to air dry. Next, the sample is placed on a hot plate at 70 °C to better adhere the film to the new substrate. The sample is then submerged in acetone on a hot plate set to 45 °C to remove the PMMA. The wet transfer processes is then completed, and the $Al_2O_3$ layer is deposited using a Cambridge NanotechS200 ALD system at 150 °C. The electrical contacts (10 nm of Ti followed by 40 nm of Au) are deposited using an E-beam evaporator (Lesker PVD75 E-beam Evaporator).

*Simulations*

Simulations were performed using a python script that runs a transfer matrix method (TMM) simulation[55]. The refractive index of each layer was measured using ellipsometry as well as the thicknesses of the $SiO_2$, $Al_2O_3$, and $WS_2$ layers.

*Spectroscopic ellipsometry*

The gate-dependent refractive index and phase modulation was measured using a J.A. Woolam M-2000 ellipsometer with a focusing lens to produce a spot size ~100 μm. The results were fitted to a multi-Lorentz oscillator model (See Supporting Information for details on the fitting process). As the M-2000 cannot produce normalized reflectance curves, a J.A. Woolam W-VASE ellipsometer was used to measure the angled reflectance without an applied voltage, and this spectrum was used to normalize the gate-dependent reflectance spectra of the M-2000.

## Acknowledgements

D. J. and J. L. acknowledge primary support for this work from the Office of Naval Research Metamaterials Program (N00014-23-1-203) and partial support from Northrop Grumman. D.J. also acknowledges partial support from the Asian Office of Aerospace Research and Development (AOARD) of the Air Force Office of Scientific Research (AFOSR) from grant no. FA2386-21-1-4063. The authors would like to thank Zahra Fakhraai for her support with the ellipsometer. The MOCVD $WS_2$ monolayer samples were grown in the 2D Crystal Consortium Materials Innovation Platform (2DCC-MIP) facility at Penn State which is supported by the National Science Foundation under NSF cooperative agreement NSF DMR-2039351. C.Y.C acknowledges support from the NSF Graduate Research Fellowship Program (NSF GRFP, DGE-1845298). Y.J.L. acknowledges financial support from the National Science and Technology Council, Taiwan (Grant No. NSTC-110-2124-M-001-008-MY3).

# Supporting Information:

# > 2π Phase Modulation using Exciton-Polaritons in a Two-Dimensional Superlattice

Jason Lynch[1], Pawan Kumar[1,2], Chen Chen[3], Nicholas Trainor[3,4], Shalina Kumari[3,4], Tzu-Yu Peng[5], Cindy Yueli Chen[6], Yu-Jung Lu[5], Joan Redwing[3,4], Deep Jariwala[1,*]

[1] Electrical and Systems Engineering, University of Pennsylvania, Philadelphia, PA 19104

[2] Inter-University Microelectronics Center, Leuven 3001, Belgium

[3] 2D Crystal Consortium Materials Innovation Platform, Materials Research Institute, Penn State University, University Park, PA, USA

[4] Materials Science and Engineering, Penn State University, University Park, PA, USA

[5] Research Centre for Applied Sciences, Academia Sinica, Taipei, Taiwan

[6] Department of Chemistry, University of Pennsylvania, Philadelphia, PA 19104, USA

*Corresponding author: dmj@seas.upenn.edu

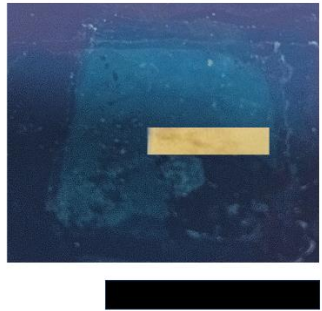

**Figure S1. Optical image of superlattice.** Optical image of the N = 3 superlattice before the external wires are connected to the electrode using an Ag paste. Scale bar is 1 cm.

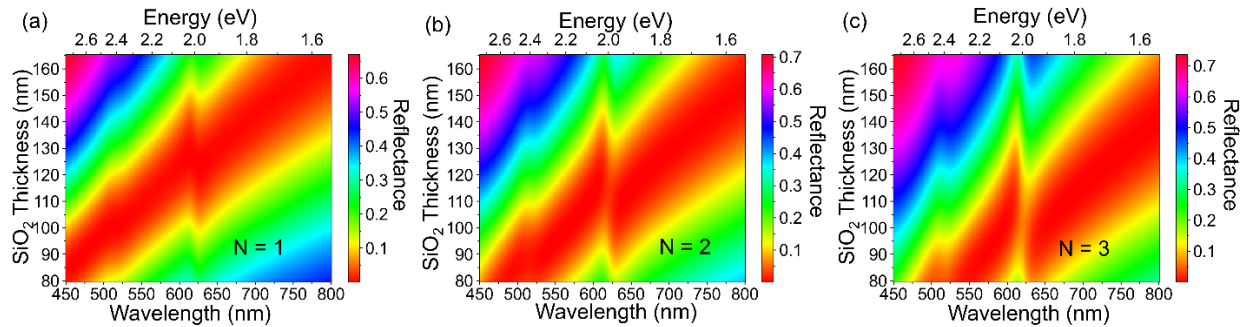

**Figure S2. Simulated layer-dependent dispersion of the WS$_2$/Al$_2$O$_3$ superlattice.** The dispersion of **(a)** N = 1, **(b)** N = 2, and **(c)** N = 3 superlattices that were calculated using the transfer matrix method (TMM). The anticrossing behavior of the exciton-polaritons is not seen in the N = 1 and 2 cases showing that they are in the weak coupling regime. The N = 3 geometry is chosen for the phase modulator since the light-matter coupling is strong enough for the hybridization of exciton and cavity modes to occur.

## *Spectroscopic Ellipsometry*

Spectroscopic ellipsometry (SE) is a highly reliable method to accurately measure the complex refractive index of films because it collects information on both the amplitude and phase of reflected light[1]. The collection of phase information makes SE more sensitive to changes in the refractive index than techniques that only collect amplitude information since the phase of light is more susceptible to change than amplitude. SE measures the ratio of the complex reflection coefficients (r) of transverse electric (TE) and transverse magnetic (TM) polarized light at oblique angles. The relative amplitude is characterized as $\psi = \arctan\left(\frac{|r_{TM}|}{|r_{TE}|}\right)$, and the relative phase is characterized as $\Delta = \varphi_{TM} - \varphi_{TE}$ where $\varphi$ is the phase of reflected light.

The first step is in the process is to perform SE on the bare SiO$_2$/Si substrate to get the exact thickness of the layers. Next, SE is performed on the same substrate with the monolayer WS$_2$, or superlattice, placed on top. The complex permittivity ($\varepsilon = (n + ik)^2$) of the sample is extracted from SE by developing a model for the system that reproduces the same Ψ and Δ values. Since WS$_2$ is a semiconductor whose optical properties in the visible range are dominated by excitonic transitions, the model consists of a series of Lorentz oscillators:

$$\varepsilon(E) = \varepsilon_\infty + \sum_i \frac{f_i \Gamma_i E_i}{E_i^2 - E^2 - iE\Gamma} \tag{S1}$$

Where E (E$_i$) is the energy of incident light (the i$^{th}$ exciton transition), $\varepsilon_\infty$ is the background permittivity, f$_i$ (Γ$_i$) is the oscillator strength (damping factor) of the i$^{th}$ exciton transition, and the sum is over all of the relevant exciton transitions. The fit parameters are then varied until the model is found to accurately replicate the experimental data. For every iteration of the fit parameters, Ψ and Δ are calculated using a transfer matrix method and compared to their experimental values using the root-mean-squared-error ($RMSE = \sqrt{\frac{1}{2p-q} \sum_i (\psi_{mod} - \psi_{exp})^2 + (\Delta_{mod} - \Delta_{exp})^2}$ where p is the number of wavelengths measured, q is the number of fit parameters, the subscript mod (exp) denoted the models (experiments), p , and the sum is over all of the measured wavelengths.

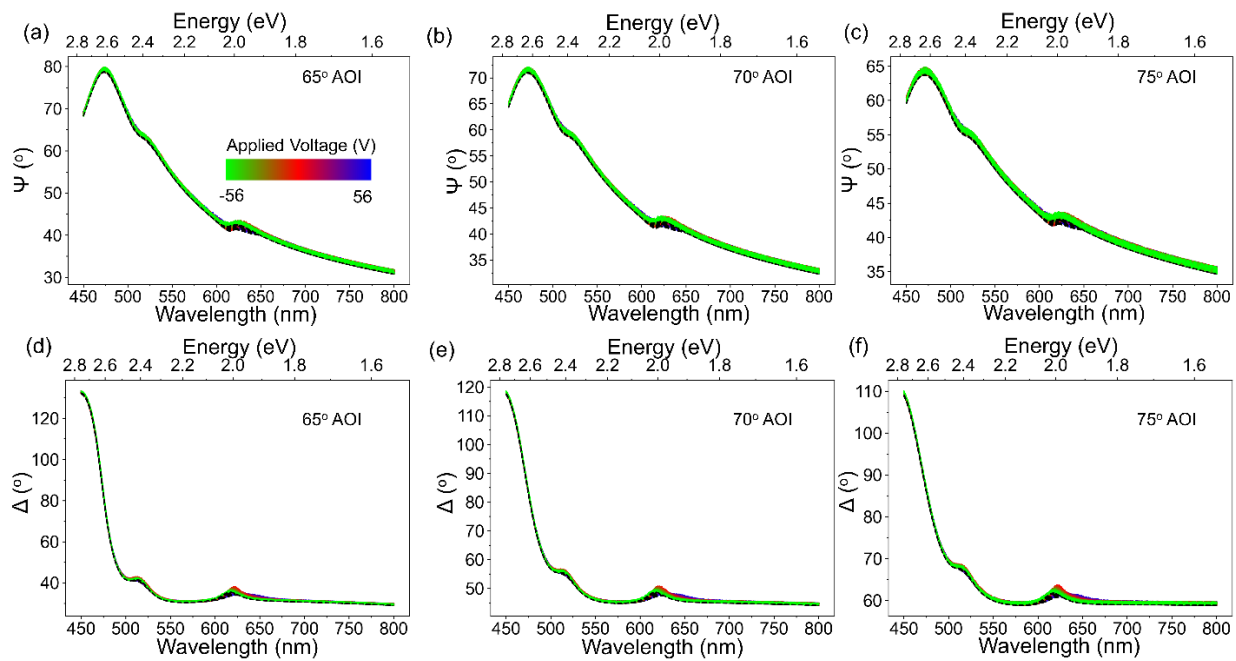

**Figure S3. Spectroscopic ellipsometry measurements of monolayer WS$_2$.** Measured (colored lines) and model (black dashed lines) values of Ψ at angles of incidence (AOI) of **(a)** 65°, **(b)** 70°, and **(c)** 75°. The corresponding values for Δ at AOIs of **(d)** 65°, **(e)** 70°, and **(f)** 75° are also shown. The experimental data is taken using an M-2000 ellipsometer. The model was fitted to the experimental data using the CompleteEase software and used the multi-Lorentz oscillator in Eq. S1. The measured sample consists of monolayer WS$_2$ on an SiO$_2$ (93 nm)/p$^{++}$-Si substrate contacted with a Ti (10 nm)/Au (40 nm) electrode.

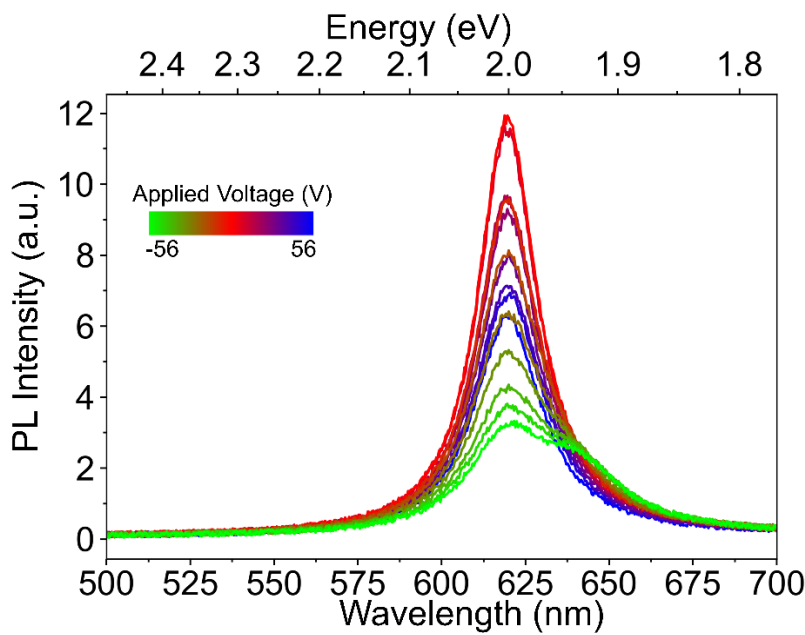

**Figure S4. Gate-tunable photoluminescence (PL) of monolayer WS$_2$.** The gate-dependent PL spectra of monolayer WS$_2$ on a SiO$_2$ (93 nm)/p$^{++}$-Si substrate. Injected charge is found to reduce the PL intensity due to a decrease in the oscillator and an increase in non-radiative recombination through scattering with free carriers. A trion shoulder is found to emerge when electrons are injected into the naturally n-type WS$_2$.

*Exciton-Polaritons*

Exciton-polaritons are part-matter, part-light quasiparticles that are the result of strong light-matter interactions that occur when the two modes are nearly resonant with one another. In the simplest model, a single exciton couples to a cavity mode through energy being exchanged between the modes. The Jaynes-Cummings Hamiltonian captures this by being the sum of the bare exciton ($H_x = E_x \sigma_z$ where $\sigma_z$ is exciton inversion operator), bare cavity ($H_c = E_c a^\dagger a$ where a ($a^\dagger$) is the photon annihilation (creation) operator), and exchange interaction ($H_{int} = g(a\sigma_+ + a^\dagger \sigma_-)$ where g is the coupling parameter $\sigma_+$ ($\sigma_-$) is the raising (lowering) operator of the exciton)[2]. As H$_{int}$ consists of one term where an exciton is excited and a photon is annihilated and another term where an exciton relaxes and a photon is created, it can be interpreted as energy flowing between the two unperturbed states. Since H$_{int}$ is proportional to g, the energy transfer rate between these is characterized by g. In matrix form, the Hamiltonian is:

$$H = \begin{pmatrix} E_x + i\Gamma_x & g \\ g & E_c(t_{SiO2}) + i\Gamma_c \end{pmatrix}$$

The eigenenergies of this Hamiltonian correspond to upper and lower polaritons, and it depends on all of the parameters present:

$$E_\pm = \frac{E_x + E_c}{2} + i\frac{\Gamma_x + \Gamma_c}{2} \pm \sqrt{4g^2 - (\Gamma_x - \Gamma_c)^2}$$

However, when a third state is present, the trion state in this case, the Hamiltonian becomes more complicated as the photon interacts with both the exciton and the trion.

$$H = \begin{pmatrix} E_x + i\Gamma_x & 0 & g_x \\ 0 & E_t + i\Gamma_t & g_t \\ g_x & g_t & E_c(t_{SiO2}) + i\Gamma_c \end{pmatrix}$$

The 3 x 3 matrix produces three eigenenergies which are labelled the upper, middle, and lower polariton depending on their energy. However, in the case where the exciton and trion peaks are not fully resolved from one another, as is in our superlattice, two of the polariton peaks become convoluted into one.

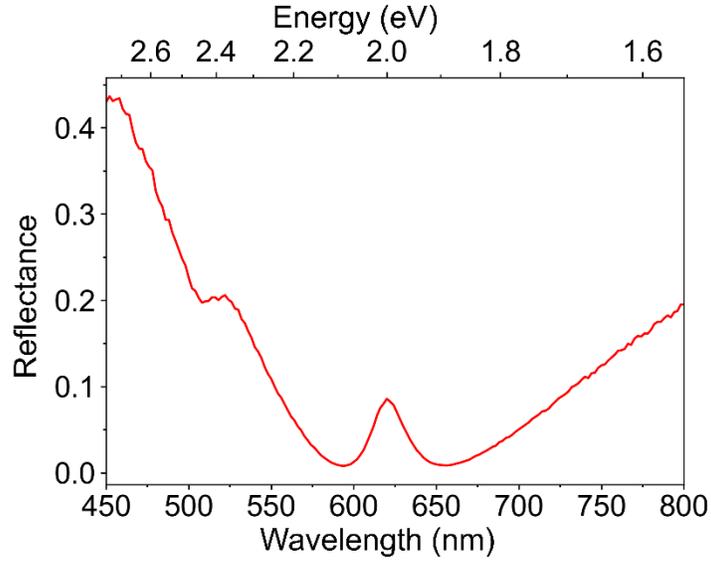

**Figure S5. Experimental reflectance spectrum of the N = 3 superlattice.** The reflectance spectra of the N = 3 superlattice for TE polarized light at an AOI of 55°. The spectrum was measured using a Woolam W-Vase ellipsometer instead of the M-2000 ellipsometer so that the reflectance could be measured with non-arbitrary units. The spectrum shows two polariton peaks. Additionally, the exciton wavelength (615 nm) corresponds to an absorptance minimum (reflectance maximum) which shows that we are in the strong light-matter coupling regime.

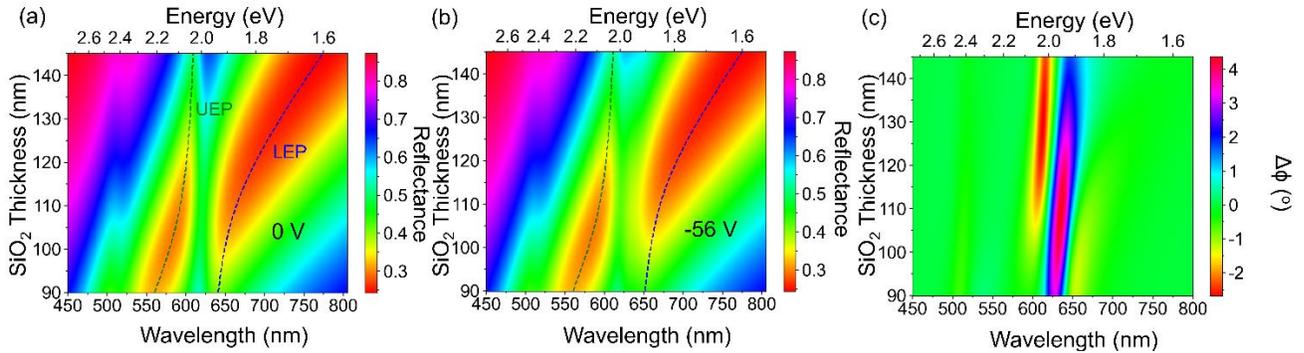

**Figure S6. Exciton-polaritons at a large AOI (80°).** The simulated dispersion with an AOI of 80° for TE polarized light incident on the superlattice and an applied voltage of **(a)** 0 V and **(b)** -56 V. The exciton coupling parameter without an applied voltage is found to be 96 meV by fitting it to the three-coupled-oscillator model. The dispersion is not very susceptible to an applied voltage as seen in the similarity in the dispersion of both applied voltages. **(c)** The simulated phase modulation of superlattice at an 80° AOI. The maximum phase modulation is < 5° showing that large AOIs are not ideal for phase modulation despite the increased coupling parameter.

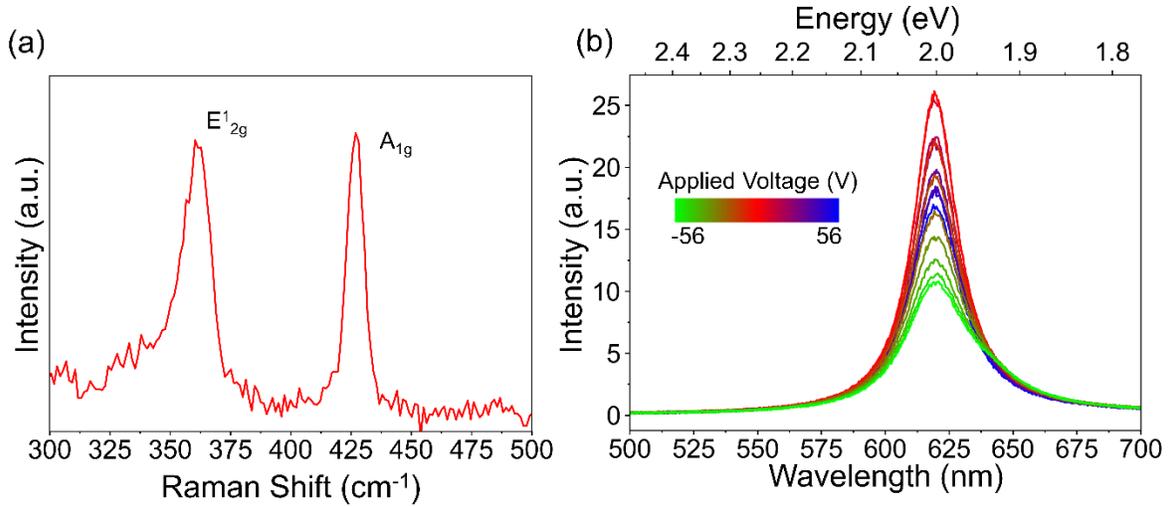

**Figure S7. Raman and gate-tunable PL spectroscopy of the N = 3 superlattice.** **(a)** Raman spectrum of the N = 3 superlattice using an excitation wavelength of 633 nm. The location of the $E^1_{2g}$ and $A_{1g}$ peaks correspond to the literature values[3] of monolayer $WS_2$ showing its high-quality after stacking. **(b)** The gate-dependent PL spectra of the N = 3 superlattice. The trion appears as a shoulder at longer wavelengths. It cannot be seen as clearly as in Figure S3 because only one of out the three layers is being modulated weakening the signal.

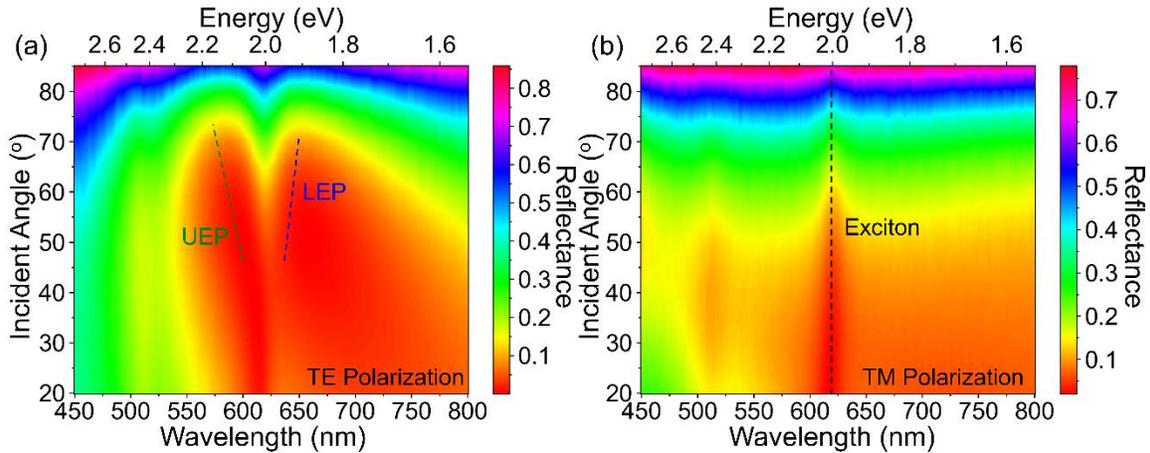

**Figure S8. Angle-dependent reflectance of the N = 3 superlattice.** The angle-dependent reflectance spectra for **(a)** TE and **(b)** TM polarized light. The spectra were measured using a Woolam W-Vase ellipsometer instead of the M-2000 ellipsometer so that the reflectance could be measured with non-arbitrary units. TE polarized light is found to produce exciton polaritons for AOIs > 50° while TM polarized light only shows excitonic effects. Both spectra were measured at steps of 5° for AOI and 2 nm for the wavelength, and the data has been smoothed using a 2D interpolation for clarity.

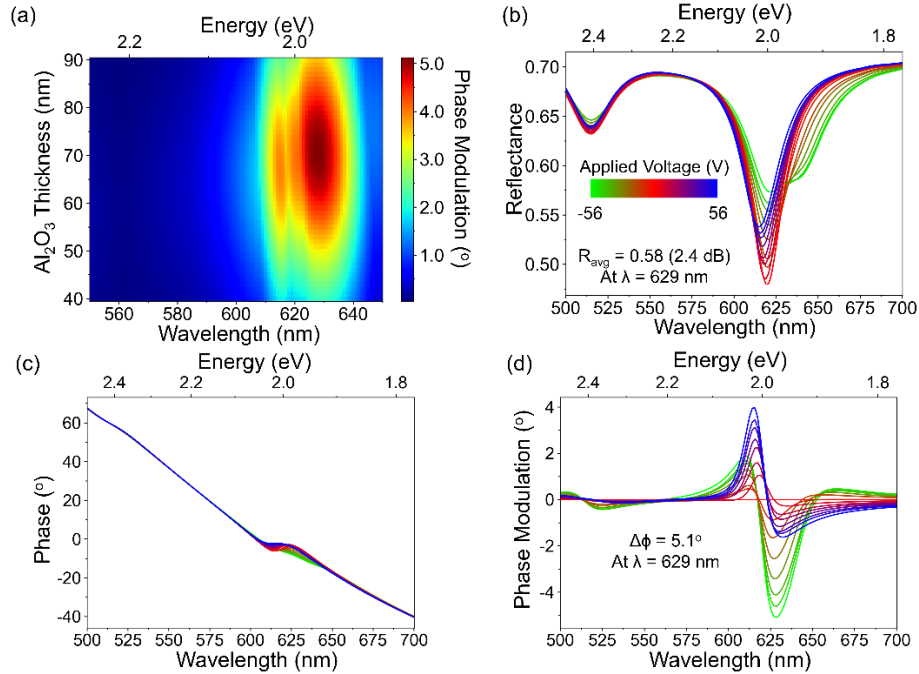

**Figure S9. Simulated performance of an exciton-trion phase modulator of WS$_2$ on Al$_2$O$_3$/Al. (a)** Optimization to maximize the phase modulation of the WS$_2$/Al$_2$O$_3$/Al system by varying the Al$_2$O$_3$ process. A thickness of 68 nm is found to produce 5.1° of phase modulation. The gate-dependent **(b)** reflectance, **(c)** phase of reflected light, and **(d)** phase modulation of the optimized system. 5.1° of phase modulation is achieved with 2.4 dB of loss giving a FoM of 2.1 °/dB. Previous work has shown a figure of merit of 51 °/dB being achieved in monolayer WS$_2$ on an Al$_2$O$_3$/Al substrate[4]. However, this work used higher-quality, exfoliated WS$_2$ flakes, and used an ion gel to inject > 4 times the number of free carriers than our work. Therefore, this simulation allows for a more direct comparison of the two systems, and our exciton-polariton modulation method increases the FoM by a factor of nearly 10 when the WS$_2$ modulation is normalized. Note that the voltages refer to the voltages we applied to the WS$_2$/SiO$_2$ (93 nm)/Si system not WS$_2$/Al$_2$O$_3$/Al.